\documentclass[12pt]{article}
\usepackage{graphicx}
\usepackage{amsmath}
\usepackage{amssymb}
\def\beq{\begin{equation}}
\def\eeq{\end{equation}}
\def\bea{\begin{eqnarray}}
\def\eea{\end{eqnarray}}
\def\ve{\vert}
\def\vel{\left|}
\def\ver{\right|}
\def\nnb{\nonumber}
\def\ga{\left(}
\def\dr{\right)}
\def\aga{\left\{}
\def\adr{\right\}}
\def\rar{\rightarrow}
\def\nnb{\nonumber}
\def\la{\langle}
\def\ra{\rangle}
\def\ba{\begin{array}}
\def\ea{\end{array}}
\def\bea{\begin{eqnarray}}
\def\eea{\end{eqnarray}}

\def\ds{\displaystyle}
\def\ve{\vert}
\def\vel{\left|}
\def\ver{\right|}
\def\nnb{\nonumber}
\def\ga{\left(}
\def\dr{\right)}
\def\aga{\left\{}
\def\adr{\right\}}
\def\rar{\rightarrow}
\def\nnb{\nonumber}
\def\la{\langle}
\def\ra{\rangle}
\def\lla{\left<}
\def\rra{\right>}

\begin{document}
\def\beq{\begin{equation}}
\def\eeq{\end{equation}}
\def\bea{\begin{eqnarray}}
\def\eea{\end{eqnarray}}
\def\ve{\vert}
\def\vel{\left|}
\def\ver{\right|}
\def\nnb{\nonumber}
\def\ga{\left(}
\def\dr{\right)}
\def\aga{\left\{}
\def\adr{\right\}}
\def\rar{\rightarrow}
\def\nnb{\nonumber}
\def\la{\langle}
\def\ra{\rangle}
\def\lla{\left<}
\def\rra{\right>}
\def\ba{\begin{array}}
\def\ea{\end{array}}
\def\BcDll{$B_c \rar (D_s, D) \ell^+ \ell^-$}
\def\BcDsll{$B_c \rar D_s^{*} \ell^+ \ell^-$}
\def\decay{$B_c \rar D_s (D_s^{*}) \ell^+ \ell^-$}
\def\tepm{$B \rar K \mu^+ \mu^-$}
\def\tept{$B \rar K \tau^+ \tau^-$}
\def\ds{\displaystyle}

\title{ {\Large {\bf
Effects of a Single Universal Extra Dimension in $B_c\rar (D_{s},D) \,\ell^+ \ell^-$  Decays} } }
\author
{{\small U. O. Yilmaz$^{1,2}$\thanks {e-mail: uoyilmaz@hacettepe.edu.tr} \, and\,  E. Danapinar $^2$\, }\\
{\small $1.$ Physics Engineering Department, Hacettepe University, 06800 Ankara, Turkey }\\
{\small $2.$ Physics Department, Karabuk University, 78100 Karabuk, Turkey } }
\date { }

\begin{titlepage}
\maketitle
\thispagestyle{empty}

\begin{abstract}
The rare semileptonic $B_c \rar D_{s,d}\, \ell^+ \ell^-$ decays are studied in the universal extra dimension with a single extra dimension scenario. The sensitivity of differential and total branching ratios, polarization asymmetries of final state leptons to the compactification parameter is presented, both for muon and tau decay channels. Comparing with the standard model, the obtained results indicate that there are new contributions to the physical observables. Considering the ability of available experiments, it would be useful to study these effects.
\end{abstract}

\end{titlepage}

\section{Introduction}
The decays induced by flavor changing neutral current (FCNC) $b\rar s, d$ transitions play an important role to test the standard model (SM) and also they are very sensitive to the new physics. The decays of $B_{u,d,s}$ mesons have been intensively studied, and it will make the B physics more complete if similar decays of $B_c$ meson are included. Considering the experimental facilities, it is possible to investigate $B_c$ decays with present ability of the LHC.

In the rare $B$ meson decays, new physics contributions appear through
the modification of the Wilson coefficients existing in the SM or by adding new structures in the SM effective Hamiltonian. Among the various extensions of the SM, extra dimensions are specially attractive because of including gravity and other interactions, giving hints on the hierarchy problem and a connection with string theory.

The models with universal extra dimensions (UED) allow the SM fields to propagate in all available dimensions \cite {Antoniadis90}-\cite{Hamed99}.
The extra dimensions are compactified and the compactification scale allows Kaluza-Klein (KK) partners of the SM fields in the four-dimensional theory and also KK excitations without corresponding SM partners.
Throughout the UED, a model including only a single universal extra dimesion is the Appelquist-Cheng-Dobrescu (ACD) model \cite{ACD}. The only additional free parameter with respect to the SM is the inverse of the compactification radius, $1/R$. In particle spectrum of the ACD model, there are infinite towers of KK modes and the ordinary SM particles are presented in the zero mode.

The experimental and theoretical discussions on the free parameter have been taken a significant part in the literature
and the lower bound was commonly taken as $1/R \geq 250\,GeV$ or $1/R \geq 350\, GeV$ \cite{Agashe}-\cite{uoyilmaz12}.
However, later analysis using the ATLAS and CMS data, the bounds increased and in some analysis \cite{Belanger}-\cite{Flacke13} $1/R < (700-715) \,GeV$ is excluded.
Here, we will exclude $1/R < 500 \,GeV$ \cite{Flacke14} and take the equivalent as the lower bound for the compactification radius.

The effective Hamiltonian of several FCNC processes \cite{Buras03}-\cite{Buras04},
semileptonic and radiative decays of B mesons \cite{Colangelo06-2}-\cite{Li11} and FCNC baryonic decays \cite{Aliev07}-\cite{Azizi11-2} have been investigated in the ACD model. Polarization properties of final state particles in semileptonic decays, which is an powerful tool in searching new physics beyond the SM, have also been studied widely besides the other observables in these works, e.g \cite{uoyilmaz12, Colangelo06-2, Saddique08, Li11}.

The main aim of this paper is to find the possible effects of the ACD model on some physical observables related to the \BcDll decays. We study differential decay rate, branching ratio, and polarization of final state leptons, including resonance contributions in as many as possible cases.
We analyze these observables in terms of the compactification factor and the form factors.
The form factors for \BcDll processes have been calculated using different quark models \cite{Geng2002}-\cite{Ebert10} and three-point QCD sum rules \cite{Azizi08}. In this work, we will use the form factors calculated in the constituent quark model \cite{Geng2002}.

This paper is organized as follows. In section 2, we give the
effective Hamiltonian for the quark level processes $b\rar (s, d)\, \ell^+ \ell^- $  with a brief discussion on the Wilson coefficients in the ACD model.
We derive matrix element using the form factors and calculate the decay rate in section 3.
In section 4, lepton polarizations are evaluated and the last two sections are dedicated to our numerical analysis, discussion on the obtained results and conclusion.
\section{Theoretical Framework}
The effective Hamiltonian describing the quark level $b\rar (s,d)\, \ell^+ \ell^- $ processes in the SM is given by \cite{Buras96}
\begin{align}
\label{effH}
{\cal H}_{eff} &=
            \frac{G_F \alpha}{\sqrt{2} \pi} V_{tq'}V_{tb}^\ast
    \Bigg[ C_{9}^{eff} (\bar q' \gamma_{\mu} L\, b)\, \bar \ell \gamma^\mu \ell
      + C_{10} (\bar q' \gamma_{\mu} L\, b)\, \bar \ell \gamma^\mu \gamma_5 \ell \nnb \\
      & -2C_{7}^{eff} m_b (\bar q' i \sigma_{\mu \nu} \frac {q^{\nu}} {q^2} R\, b)\,\bar \ell \gamma^\mu \ell
     \Bigg]\,,
\end{align}
where $q=p_{B_c}-p_{D_{q'}}$ is the momentum transfer and $q'=s, d$.

New physics effects in the ACD model come out by the modification of the SM Wilson coefficients appear in the above Hamiltonian. This process can be done by writing the Wilson coefficients in terms of $1/R$ dependent periodic functions the details of which can be found in \cite{Buras03}-\cite{Buras04}.
That is, $F_0(x_t)$ in the SM is generalized by $F(x_t, 1/R)$ accordingly
\bea
\label{WACD} F(x_t, 1/R) = F_0(x_t) + \sum_{n=1} ^{\infty} F_n
(x_t, x_n)
\eea
where $x_t=m_t^2/m_W^2$, $x_n=m_n^2/m_W^2$ and $m_n=n/R$. These modified Wilson coefficients can widely be found in the literature. Here, we will not discuss the details and follow the explanation given in  \cite{uoyilmaz12}.

In the ACD model, a normalization scheme independent effective coefficient $C_7^{eff}$ can be written as
\begin{align}
\label{C7eff} C_7^{eff}(\mu_b, 1/R) &=  \eta^{16/23} C_7(\mu_W, 1/R) \nnb \\
            & + \frac{8}{3} (\eta^{14/23} - \eta^{16/23}) C_8(\mu_W, 1/R) +
                    C_2(\mu_W, 1/R) \sum_{i=1} ^{8} h_i \eta^{a_i}\,.
\end{align}
The coefficient $C_9^{eff}$ has perturbative part and we will also consider the resonance contributions coming from
the conversion of the real $c \bar c$ into lepton pair. So, $C_9^{eff}$ is given by
\bea
C_9^{eff}(s', 1/R) = C_9(\mu, 1/R)\Big(1+\frac{\alpha_s(\mu)} {\pi} w(s')\Big) + Y(\mu)+C_9^{res}(\mu)\,.
\eea
For $C_9$, in the ACD model and in the naive dimensional regularization (NDR) scheme we have
\bea
\label{C9mu} C_9(\mu, 1/R)= P_0^{NDR} + \frac {Y(x_t, 1/R)} {sin^2{\theta_W}} - 4 Z(x_t, 1/R) + P_E E(x_t, 1/R)
\eea
where $P_0^{NDR}=2.60\pm 0.25$ and the last term, $P_E E(x_t, 1/R)$, is numerically negligible.

The perturbative part, coming from one-loop matrix elements of the four quark operators, is
\begin{align}
\label{EqY} Y(\mu)&= h(y,s) [ 3 C_1(\mu) + C_2(\mu) + 3
         C_3(\mu) + C_4(\mu) + 3 C_5(\mu) + C_6(\mu)] \nnb \\
    &- \frac{1}{2} h(1, s) \left( 4 C_3(\mu) + 4 C_4(\mu)
        + 3 C_5(\mu) + C_6(\mu) \right)\nnb \\
    &- \frac{1}{2} h(0,  s) \left[ C_3(\mu) + 3 C_4(\mu) \right] \nnb \\
    &+ \frac{2}{9} \left( 3 C_3(\mu) + C_4(\mu) + 3 C_5(\mu) +
            C_6(\mu) \right)\, ,
\end{align}
with $y=m_c/m_b$. The explicit forms of the functions appear above equations can be found in \cite{Buras95}-\cite{Misiak93}. The resonance contribution can be done by using a Breit-Wigner formula \cite{AAli91}
\begin{align}
C^{res}_{9}&=-\frac{3}{\alpha^2_{em}}\kappa \sum_{V_i=\psi_i}
        \frac{\pi \Gamma(V_i\rightarrow \ell^+ \ell^-)m_{V_i}}{s m_b^2 -m^{2}_{V_i}+i m_{V_i}
        \Gamma_{V_i}} \nonumber \\
    &\times  [ 3 C_1(\mu) + C_2(\mu) + 3 C_3(\mu) + C_4(\mu) + 3
         C_5(\mu) + C_6(\mu)]\,.
 \label{Yresx}
\end{align}
The normalization is fixed by the data in \cite {pdg} and $\kappa$ is taken 2.3.

The Wilson coefficient $C_{10}$ is independent of scale $\mu$ and given by
\bea
\label{C10} C_{10} = - \frac{Y(x_t, 1/R)} {sin^2{\theta_W}}\,.
\eea
\section {Matrix Elements and Decay Rate}
The matrix elements for $B_c\rar D_{q'} \ell^+ \ell^-$ can be written in terms of the invariant form factors over $B_c$ and $D_{q'}$. The parts of transition currents containing $\gamma_5$ do not contribute, so the non-vanishing matrix elements are \cite{AAli2000}
\begin{align}
\label{formfactor}
    \lla D_{q'}(p_{D_{q'}}) \vel \bar {q'}  i \sigma_{\mu \nu} q^{\nu} b \ver B_c(p_{B_c}) \rra
   & = -\frac{f_T(q^2)}{m_{B_c}+m_{D_{q'}}} \Big[(P_{B_c}+P_ {D{q'}})_{\mu} q^2 - q_{\mu} (m^2_{B_c}-m^{2}_{D_{q'}})\Big] \,, \nnb \\
    \lla D_{q'}(p_{D_{q'}}) \vel \bar {q'}  \gamma _{\mu} b \ver B_c(p_{B_c}) \rra
    &= f_{+}(q^2)(P_{B_c} + P_{D_{q'}})_{\mu} + f_{-}(q^2) q_{\mu}\, .
\end{align}
\begin{figure}[h]
\centering
\includegraphics[scale=0.60]{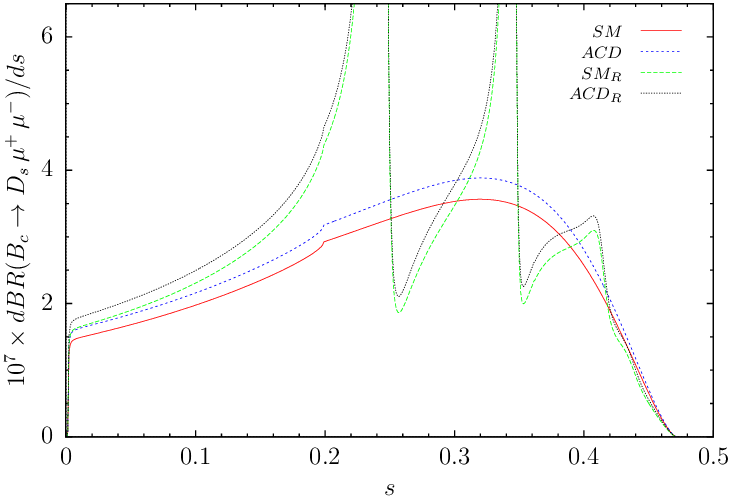}~~~
\includegraphics[scale=0.60]{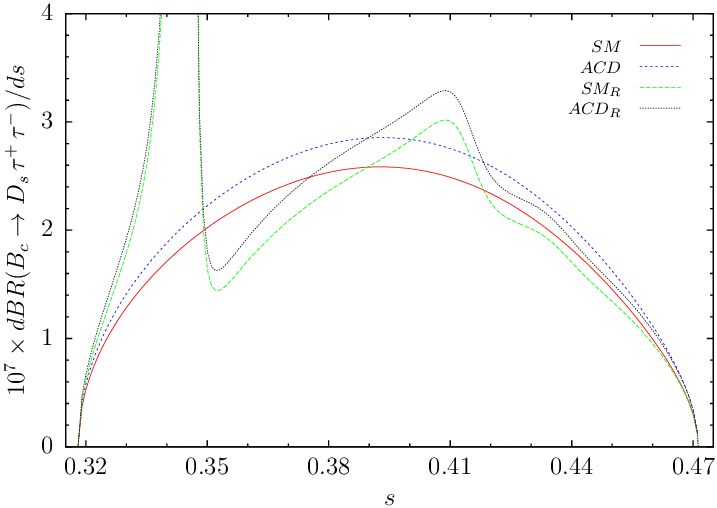}
\caption{The dependence of differential branching ratio on s with and without resonance contributions for $B_c\rar D_s \ell^+ \ell^-$ in the SM and the ACD Model for $1/R=500\,GeV$. (The subscript R represents resonance contribution.) \label{dbr-dc}}
\end{figure}
\begin{figure}[h]
\centering
\includegraphics[scale=0.60]{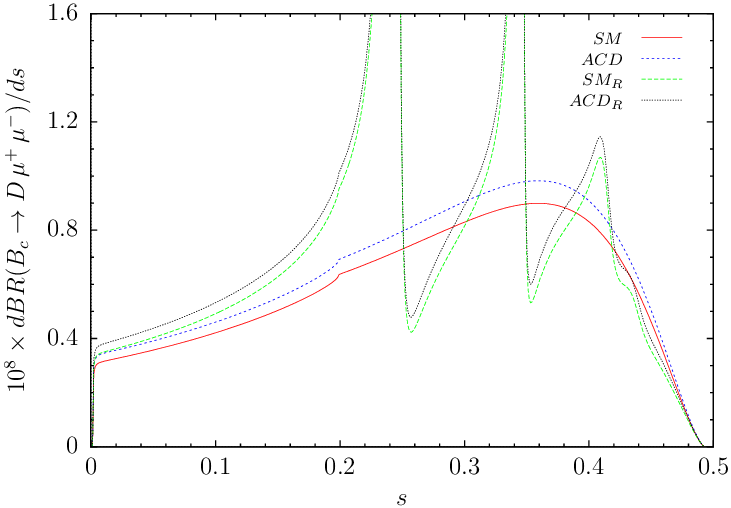}~~~
\includegraphics[scale=0.60]{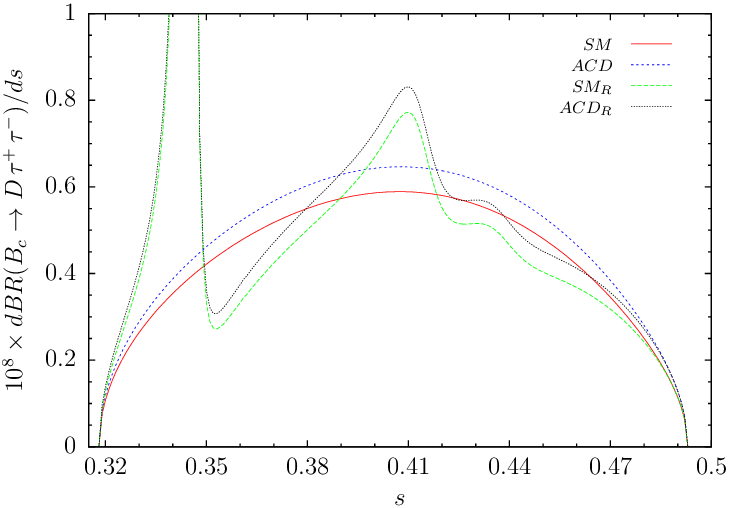}
\caption{The dependence of differential branching ratio on s with and without resonance contributions  for $B_c\rar D \ell^+ \ell^-$ in the SM and the ACD Model for $1/R=500\,GeV$. \label{dbr-dd}}
\end{figure}

The transition amplitude of the $B_c\rar D_{q'} \ell^+ \ell^-$ decays can be written
using the effective Hamiltonian and eq. (\ref {formfactor}) as
\begin{align}
 \label{bdhad}
    {\cal M}(B_c\rightarrow D_{q'} \ell^{+}\ell^{-}) &=
    \frac{G \alpha}{2 \sqrt{2} \pi} V_{tb} V_{tq'}^\ast \Bigg\{
            \bar \ell \gamma^\mu \ell \, \Big[
            A(P_{B_c} + P_{D_{q'}})_\mu +Bq_\mu \Big] \nnb \\
&+ \bar \ell \gamma^\mu \gamma_5 \ell \, \Big[C(P_{B_c} + P_{D_{q'}})_\mu +Dq_\mu \Big] \Bigg\}~,
\end{align}
with
\begin{align}
\label{bdas}
&A = C_{9}^{eff} f_{+} + \frac{2 m_{b} f_T}{m_{B_c}+m_{D_{q'}}}C_7^{eff}~, \nnb \\
&B = C_{9}^{eff} f_{-} - \frac{2 m_{b} (m^2_{B_c}-m^2_{D_{q'}}) f_T}{q^2 (m_{B_c} + m_{D_{q'}})} C_7^{eff}~, \nnb \\
&C = C_{10} f_{+}~, \nnb \\
&D = C_{10} f_{-}~.
\end{align}

Finally,  following dilepton mass spectrum is obtained by eliminating angular dependence in the double differential decay rate,
\bea
\label{bdunp} \frac{d
    \Gamma}{ds} = \frac{G^2 \alpha^2
        m_{B_c}}{2^{12} \pi^5 }
         \vel V_{tb} V_{tq'}^\ast \ver^2 \sqrt{\lambda} v \Delta_{D_{q'}}
\eea
where $s=q^2/m_{B_c}^2$, $\lambda= 1 + r^2 + s^2 -2r-2s-2rs$,
$r=m_{D_s^\ast}^2/m_{B_c}^2$, $v=\sqrt{1-{4m_\ell^2}/{s m_{B_c}^2}}$ and
\begin{align}
\label{bddelta} \Delta_{D_{q'}} &= \frac{4}{3} m^4_{B_c}(3-v^2)\lambda (\vel A \ver^2 +\vel C \ver^2)
            + 4 m^4_{B_c} s (2+r-s) (1-v^2) \vel C \ver^2 \nnb \\
& + 16 m^2_{B_c} m^2_{\ell}s \vel D \ver^2 +32 m^2_{B_c} m^2_{\ell} (1-r) Re(CD^\ast)\,.
\end{align}

\begin{figure}[h]
\centering
\includegraphics[scale=0.6]{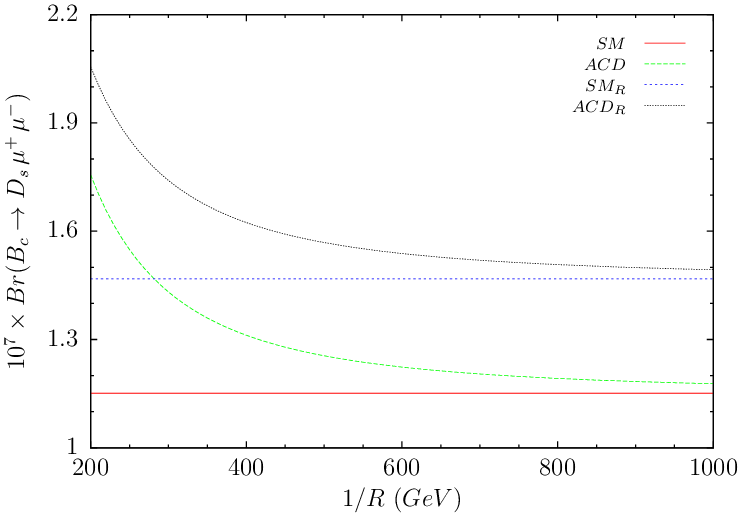}~~~
\includegraphics[scale=0.6]{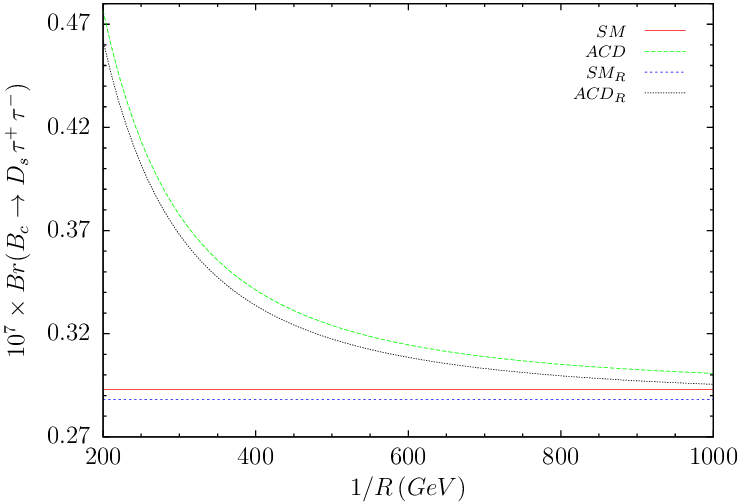}
\caption{The dependence of branching ratio on 1/R with and without resonance contributions for $B_c\rar D_s \ell^+ \ell^-$. (The subscript R represents resonance contribution.) \label{br-dc}}
\end{figure}
\begin{figure}[h]
\centering
\includegraphics[scale=0.6]{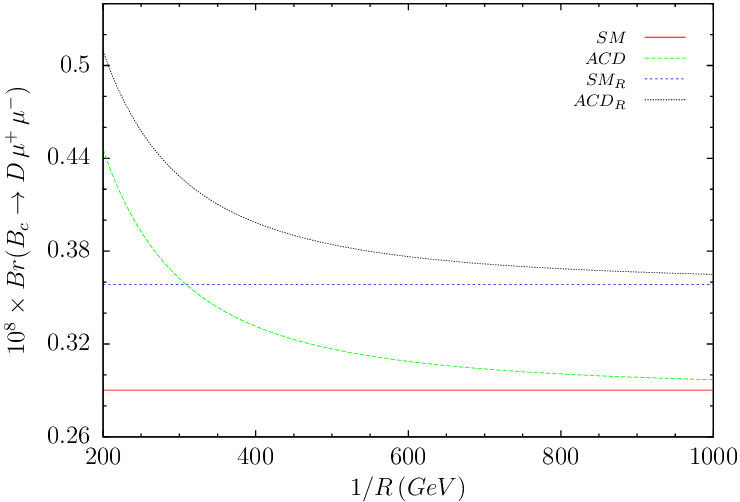}~~~
\includegraphics[scale=0.6]{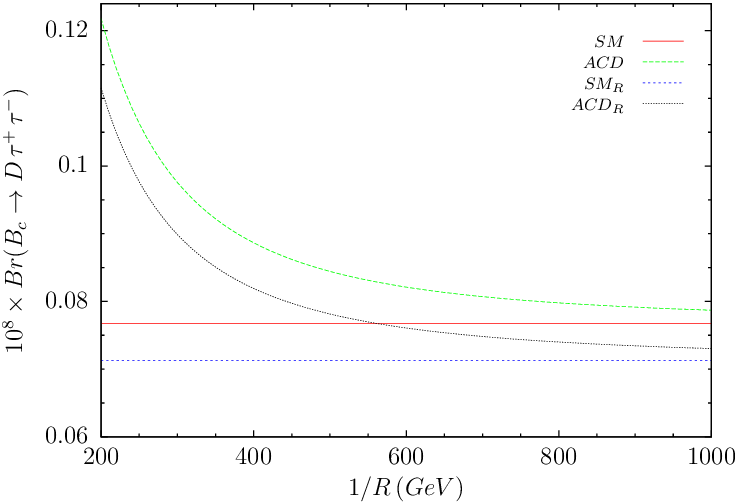}
\caption{The dependence of branching ratio on 1/R with and without resonance contributions for $B_c\rar D \ell^+ \ell^-$. \label{br-dd}}
\end{figure}
\section {Lepton Polarization Asymmetries}
Studying polarization asymmetries of final state leptons is an useful way of searching new physics.
Therefore, we will discuss the possible effects of the ACD model in the lepton polarization. Using the convention in \cite{Fukae2000}-\cite{Kruger96}, we define the orthogonal unit vectors $S_i^{-}$  ($i=L, T, N$) for the longitudinal,  transverse and normal polarizations
in the rest frame of $\ell^-$ as
\begin{align}
\label{pol}
&S_L^{-} \equiv (0,\vec{e}_L) =
    \ga 0,\frac{\vec{p}_{\ell}}{\vel \vec{p}_{\ell} \ver} \dr, \nnb \\
&S_T^{-} \equiv (0,\vec{e}_T) =
    \ga 0, \vec{e}_N \times \vec{e}_L \dr, \nnb \\
&S_N^{-} \equiv (0,\vec{e}_N) =
    \ga 0,\frac{\vec{p}_{D_{q'}} \times \vec{p}_{\ell}}
    {\vel \vec{p}_{D_{q'}} \times \vec{p}_{\ell} \ver} \dr,
\end{align}
where $\vec{p}_{\ell}$ and $\vec{p}_{D_{q'}}$ are the three momenta of $\ell^-$ and
$D_{q'}$ meson in the center of mass (CM) frame of final state leptons, respectively.
The longitudinal unit vector $S_L^-$ is boosted by Lorentz transformation,
\bea
\label{bs}
S^{-\mu}_{L,\, CM} = \ga \frac{\vel \vec{p}_{\ell} \ver}{m_\ell},
    \frac{E_\ell \,\vec{p}_{\ell}}{m_\ell \vel \vec{p}_{\ell} \ver} \dr,
\eea
while vectors of perpendicular directions remain unchanged under the Lorentz boost.

The differential decay rate of $B_c \rar D_{q'} \ell^+ \ell^-$ for
any spin direction $\vec{n}^{-}$ of the $\ell^{-}$
can be written in the following form
\bea
\label{ddr}
\frac{d\Gamma(\vec{n}^{-})}{ds} = \frac{1}{2}
\ga \frac{d\Gamma}{ds}\dr_0
\Bigg[ 1 + \Bigg( P_L^{} \vec{e}_L^{\,-} + P_N^{-}
\vec{e}_N^{\,-} + P_T^{-} \vec{e}_T^{\,-} \Bigg) \cdot
\vec{n}^{-} \Bigg]~,
\eea
where $(d \Gamma / ds)_0$ corresponds to the unpolarized decay
rate, the explicit form of which is given in eq. (\ref{bddelta}).

The polarizations $P^{-}_L$,  $P^{-}_T$ and $P^{-}_N$ in eq. (\ref{ddr}) are defined by the equation
\bea P_i^{-}(s) = \frac{\ds{{d \Gamma}
                   ({\bf{n}}^{-}={\bf{e}}_i^{\,-})/{ds} -
                   {d \Gamma}
                   ({\bf{n}}^{-}=-{\bf{e}}_i^{\,-})/{ds}}}
             {\ds{{d \Gamma}
             ({\bf{n}}^{-}={\bf{e}}_i^{\,-})/{ds} +
             {d \Gamma}
             ({\bf{n}}^{-}=-{\bf{e}}_i^{\,-})/{ds}}}\,. \nnb
\eea
Here, $P^{-}_L$ and $P^{-}_T$ represent the longitudinal and transversal asymmetries, respectively, of the charged lepton $\ell^{-}$ in the
decay plane, and $P^{-}_N$ is the normal component to both of them.
\begin{figure}[h]
\centering
\includegraphics[scale=0.6]{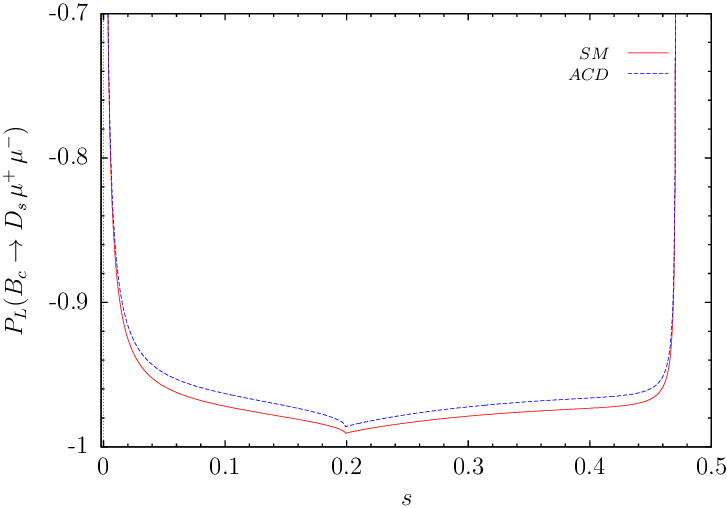}~~~
\includegraphics[scale=0.6]{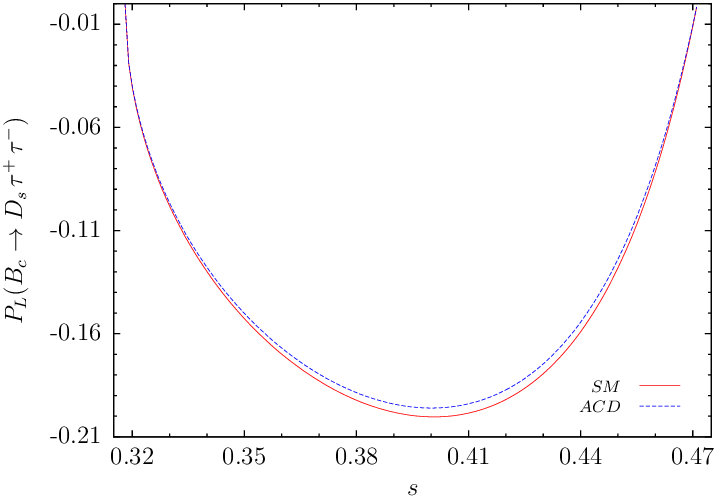}
\caption{The dependence of longitudinal polarization on s without resonance contributions for $B_c\rar D_s \ell^+ \ell^-$. \label{plong-dc}}
\end{figure}
\begin{figure}[h]
\centering
\includegraphics[scale=0.6]{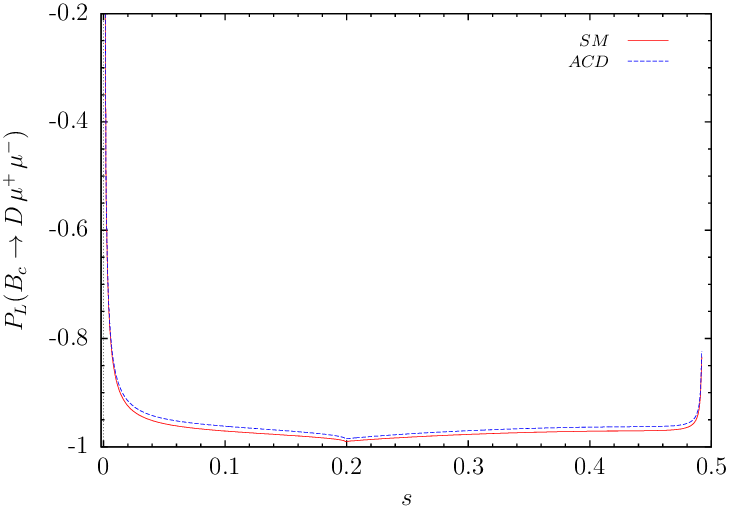}~~~
\includegraphics[scale=0.6]{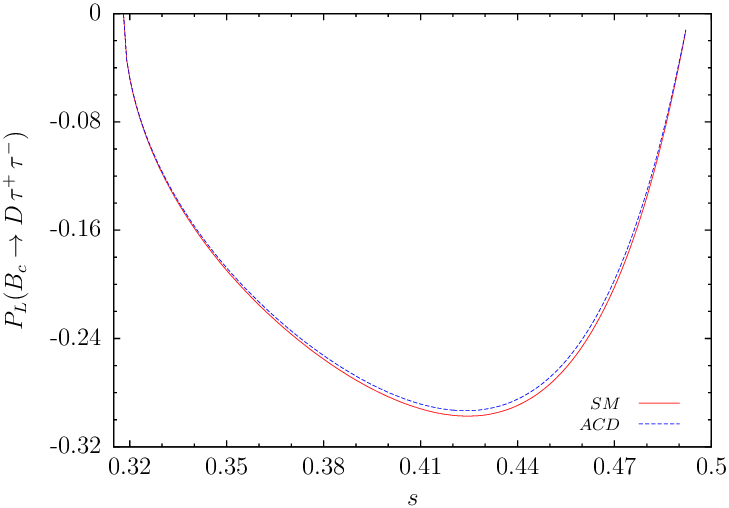}
\caption{The dependence of longitudinal polarization on s without resonance contributions for $B_c\rar D \ell^+ \ell^-$. \label{plong-dd}}
\end{figure}

The straightforward calculations yield the explicit form of the longitudinal polarization for $B_c \rar D_{s,d}\, \ell^+ \ell^-$ as
\bea
\label{bdlong}
P^-_L=
 \frac{16} {3 \Delta}m^4_{B_c}v \lambda Re[AC^\ast]\,,
\eea
and the transversal polarization is given by
\bea
\label{bdtrans}
P^-_T= \frac{4 m^3_{B_c} m_\ell \pi \sqrt{s \lambda}}{\Delta} \Bigg[\frac{(r-1)}{s} Re[AC^\ast]+Re[AD^\ast] \Bigg].
\eea
\begin{figure}[h]
\centering
\includegraphics[scale=0.6]{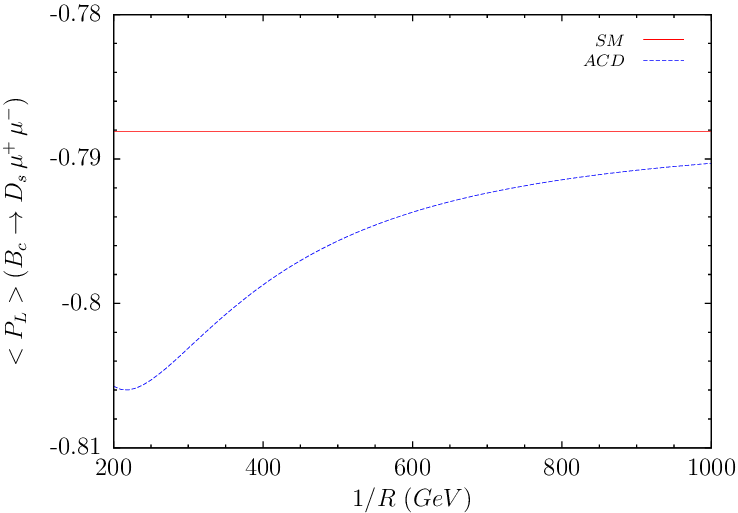}~~~
\includegraphics[scale=0.6]{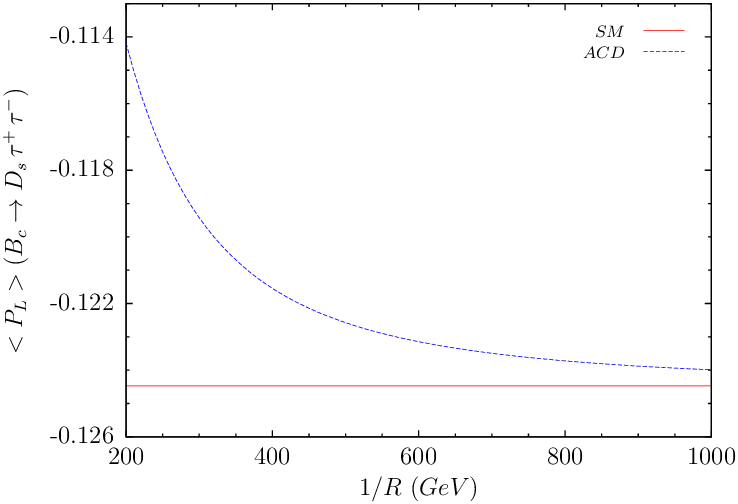}
\caption{The dependence of longitudinal polarization on $1/R$ with resonance contributions for $B_c\rar D_s \ell^+ \ell^-$. \label{plong-av-dc}}
\end{figure}
\begin{figure}[h]
\centering
\includegraphics[scale=0.6]{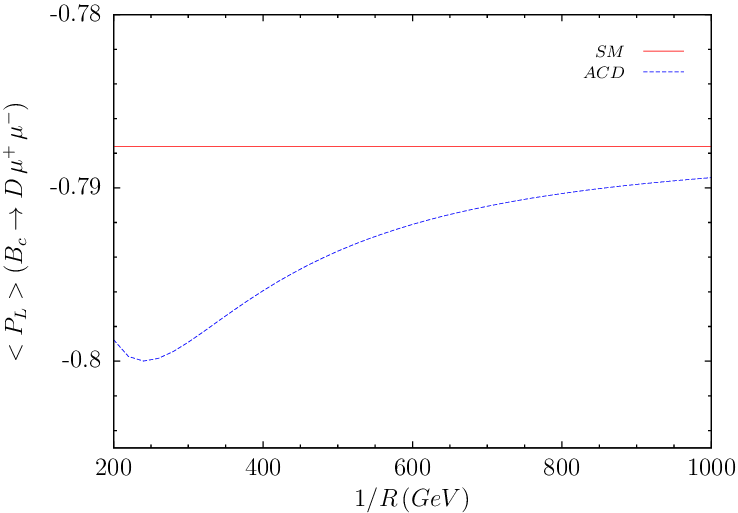}~~~
\includegraphics[scale=0.6]{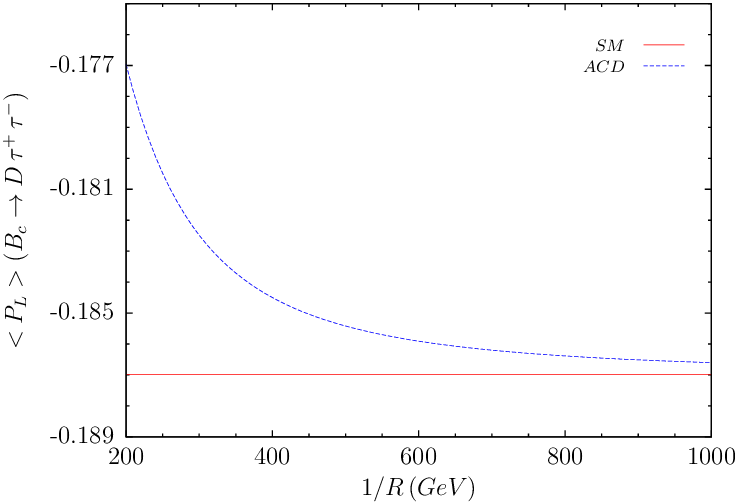}
\caption{The dependence of longitudinal polarization on $1/R$ with resonance contributions for $B_c\rar D \ell^+ \ell^-$. \label{plong-av-dd}}
\end{figure}
The normal component of polarization is zero so we have not stated its explicit form here.
\section{Numerical Analysis}
In this section, we will introduce numerical analysis of physical observables. Some of the input parameters used in this work are $m_{B_c}=6.28 \, GeV$, $m_{D_{s}}=1.968 \,GeV$, $m_{D}=1.870 \,GeV$,
$m_b =4.8 \, GeV$, $m_{\mu} =0.105 \, GeV$, $m_{\tau} =1.77 \, GeV$, $|V_{tb} V^*_{ts}|=0.041$,
$|V_{tb} V^*_{td}|=0.008$,
$G_F=1.17 \times 10^{-5}\, GeV^{-2}$ and $\tau_{B_{c}}=0.46 \times 10^{-12} \, s$ \cite {pdg}.

To make numerical predictions, we also need the explicit forms
of the form factors $f_+, f_-$ and $f_T$.  In our analysis, we used the
results of \cite{Geng2002}, calculated in the constituent quark model and $q^2$ parametrization is given by
\begin{eqnarray}
F(q^2) = \frac{F(0)} {1-a (q^2/m^2_{B_c}) + b (q^2/m^2_{B_c})^2}~, \nnb
\end{eqnarray}
where the values of parameters $F(0)$, $a$ and $b$ for the $B_c \rar (D_s, D)$ decays are listed in Table \ref{form-table}.

\begin{figure}[h]
\centering
\includegraphics[scale=0.6]{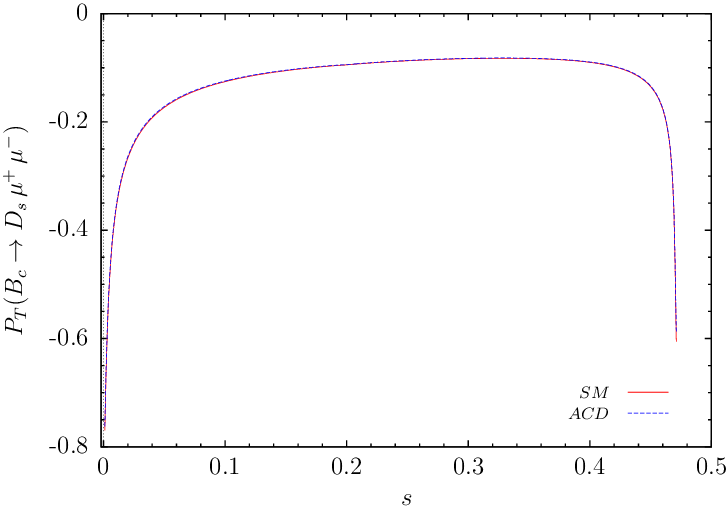}~~~
\includegraphics[scale=0.6]{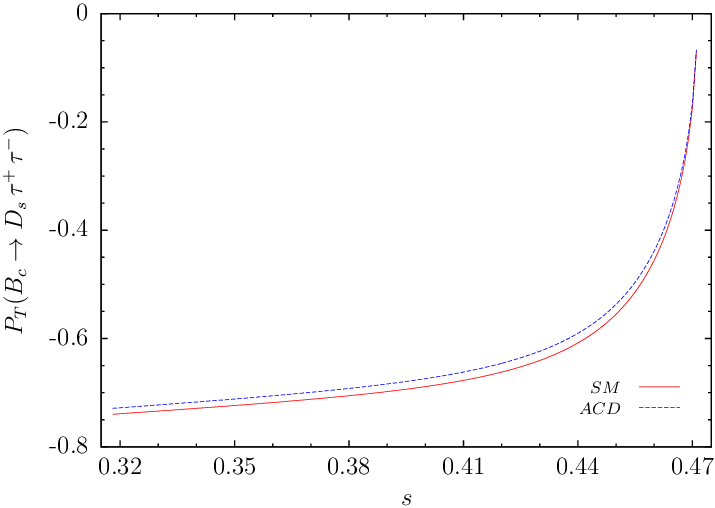}
\caption{The dependence of transversal polarization on s without resonance contributions for $B_c\rar D_s \ell^+ \ell^-$. \label{ptrans-dc}}
\end{figure}
\begin{figure}[h]
\centering
\includegraphics[scale=0.6]{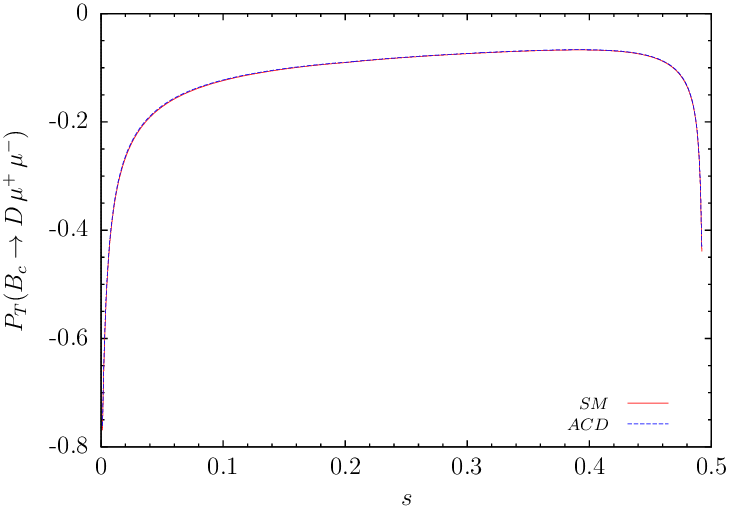}~~~
\includegraphics[scale=0.6]{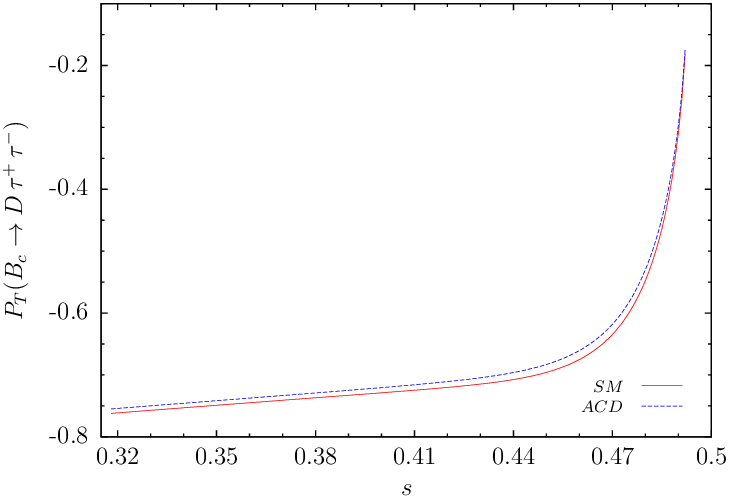}
\caption{The dependence of transversal polarization on s without resonance contributions for $B_c\rar D \ell^+ \ell^-$. \label{ptrans-dd}}
\end{figure}

In this work, we also took the long-distance contributions into account. While doing this, to minimize the hadronic uncertainties we introduce some cuts around ${J/ \psi}$ and $\psi(2s)$ resonances as discussed in \cite{uoyilmaz12}.

In the analysis, first the differential branching ratios are calculated with and without resonance contributions and s dependence for the SM and $1/R=500 \,GeV$ are presented in Figures \ref{dbr-dc} and \ref{dbr-dd} for $B_c\rar (D_{s} , D) \ell^+ \ell^-$. One can notice the change in the differential decay rate and difference between the SM results and new effects in the figures. The maximum deviation is around $s=0.32 \,(0.39)$ in Figure \ref{dbr-dc} and $s=0.36\,(0.40)$ in Figure \ref{dbr-dd} for $\mu\,(\tau)$. The deviation is $\sim 10 \%$ and less for $1/R> 500\, GeV$.
Considering the resonance effects, the differential decay rates also differ from their SM values.

To introduce the contributions of the ACD model on the branching ratio, we present $1/R$ dependent ratios with and without resonance cases in Figures \ref{br-dc} and \ref{br-dd}.
The common feature is that as $1/R$ increases, the branching ratios approach to their SM values and
vary in the following ranges for $1/R \geq  500\, GeV$,
\begin{align}
&Br(B_c\rar D_s \mu^+ \mu^-)= (1.151-1.255)\times 10^{-7} \nnb \\
&Br(B_c\rar D_s \tau^+ \tau^-)=(0.293-0.324)\times 10^{-7} \nnb \\
&Br(B_c\rar D~ \mu^+ \mu^-)= (0.290-0.317)\times 10^{-8} \nnb \\
&Br(B_c\rar D~ \tau^+ \tau^-)=(0.077-0.078) \times 10^{-8} \nnb.
\end{align}

Here, the first value in any branching ratios above is corresponding to the SM,  while the second one is for $1/R=500\,GeV$, without resonance contributions.
A similar behavior is valid for resonance case which can be followed by the figures.

Adding the uncertainty on the form factors may influence the contribution range of the ACD model.
However, the variation of the branching ratios, calculated with the central values of form factors, in the ACD model with the SM values, can be considered as a signal of new physics.
\begin{figure}[h]
\centering
\includegraphics[scale=0.6]{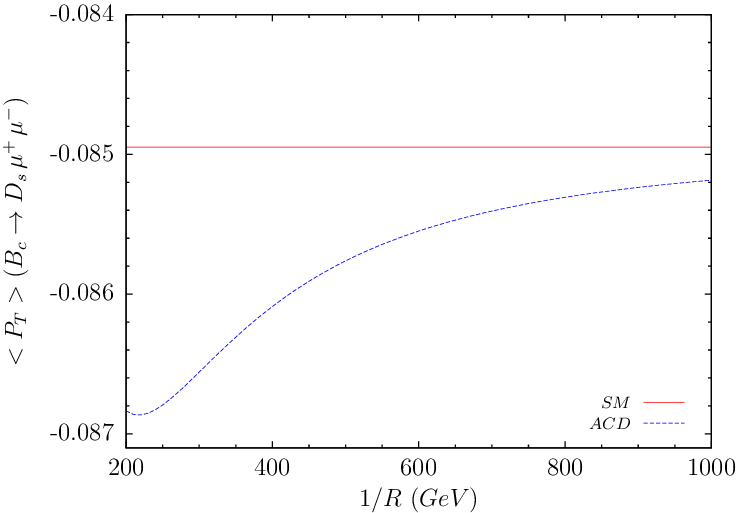}~~~
\includegraphics[scale=0.6]{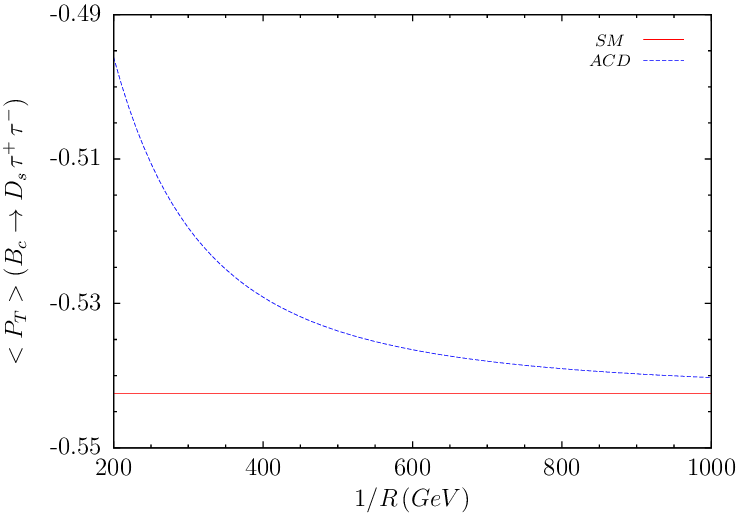}
\caption{The dependence of transversal polarization on $1/R$ with resonance contributions for $B_c\rar D_s \ell^+ \ell^-$. \label{ptrans-av-dc}}
\end{figure}
\begin{figure}[h]
\centering
\includegraphics[scale=0.6]{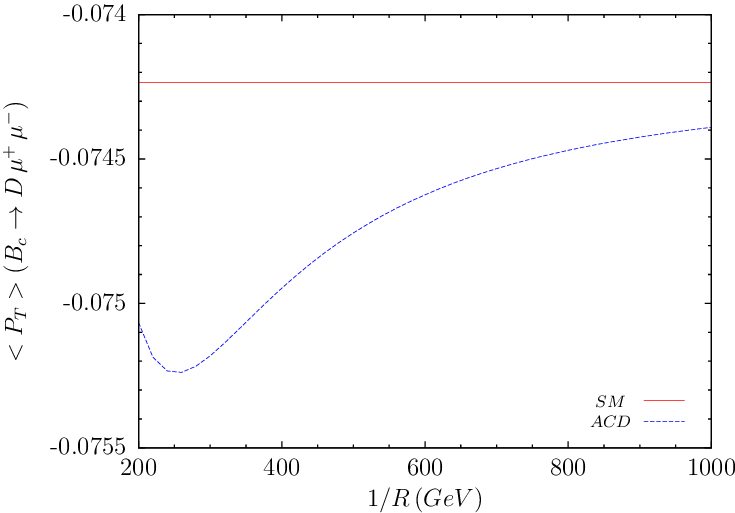}~~~
\includegraphics[scale=0.6]{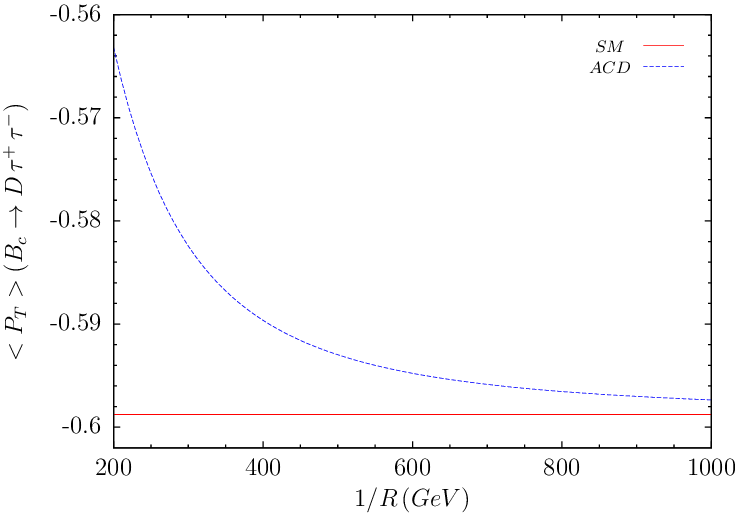}
\caption{ The dependence of transversal polarization on $1/R$ with resonance contributions for $B_c\rar D \ell^+ \ell^-$. \label{ptrans-av-dd}}
\end{figure}

The polarization properties of final state leptons give useful clues for new physics. Hence,
the dependence of longitudinal polarization on s without resonance contributions are given by Figures  \ref{plong-dc} and \ref{plong-dd}. The longitudinal polarization differ from the SM values slightly. The maximum deviation can be found less than $5\%$.

In order to clarify the dependence on $1/R$, we eliminate the dependence of the lepton polarizations on $s$, by considering the averaged forms over the allowed kinematical region as
\bea
\label{av} \lla P_i \rra =
        \frac{\ds \int_{(2 m_\ell/m_{B_c})^2}^{(1-m_{D_{q'}}/m_{B_c})^2}
            P_i \frac{d{\cal B}}{ds} ds} {\ds \int_{(2
        m_\ell/m_{B_c})^2}^{(1-m_{D_{q'}}/m_{B_c})^2}
 \frac{d{\cal B}}{ds} ds}.
\eea
\begin{table}[h]
\renewcommand{\arraystretch}{1.5}
\addtolength{\arraycolsep}{3pt}
\caption{$B_c \rar D_{s,d}$ decays form factors calculated in the constitute quark model. \label{form-table}}
$$
\begin{array}{lccc}
\hline
\hline
B_c \rar D_s \ell^+ \ell^- &   F(0) & a & b \\ \hline
f_+ &            ~~\, 0.165    & -3.40   & 3.21  \\
f_- &            -0.186    & -3.51   & 3.38  \\
f_T &            -0.258    & -3.41   & 3.30  \\
\hline
B_c \rar D \ell^+ \ell^- & F(0)   & a & b \\ \hline
f_+ &        ~~\,0.126         & -3.35 &  3.03  \\
f_- &           -0.141         & -3.63 &  3.55  \\
f_T &           -0.199         & -3.52 &   3.38   \\ \hline \hline
\end{array}
$$
\renewcommand{\arraystretch}{1}
\addtolength{\arraycolsep}{-3pt}
\end{table}
The $1/R$ dependant average longitudinal polarizations are given in Figures \ref{plong-av-dc} and \ref{plong-av-dd}.
As it can be seen from the figures, the maximum deviation is $2\%$ for $\mu$ channels and  $3\%$  for $B_c\rar (D_s, D)\tau^+ \tau^-$,
respectively, at $1/R=500\, GeV$.

The variation of transversal polarization with respect to $s$ are given by Figures \ref{ptrans-dc} and \ref{ptrans-dd}. In $\mu$ channels the difference is negligible. In $\tau$ channels up to $s\sim 0.46\, (0.48)$ for $B_c\rar D_s (D) \ell^+ \ell^-$ decays, respectively, the effects of the UED can be followed.
Although the deviation with the SM values are very small
Finally, the average transversal polarization can be observed by Figures \ref{ptrans-av-dc} and \ref{ptrans-av-dd}.

\section{Conclusion}
In this work, we have studied $B_c\rar D_s (D) \ell^+ \ell^-$ in the framework of a single universal extra dimension
and calculated the contributions to some physical observables.
\\
As an overall conclusion, as $1/R\rar 200\,GeV$ the physical values differ from the SM results. However, considering the lower bound on the compactification
factor which is above $500\,GeV$, there still appears acceptable difference on the differential and integrated branching ratios comparing with the SM results.

The polarization effects in $\mu$ channels are either irrelevant nor negligible while in $\tau$ channels some small effects are obtained.
Under the discussion in this work, studying these decays experimentally can be useful for understanding new physics.
\\
\\
\large {\textbf{Competing Interests}}\\
The authors declare that there is no conflict of interest 
regarding the publication of this paper.


\begin{thebibliography}{99}
%
\bibitem{Antoniadis90} I. Antoniadis,"A Possible new dimension at a few TeV",  \emph{Physics Letters  B}, vol. 246, no. 3-4, pp. 377-384, 1990.

\bibitem{Antoniadis98} I. Antoniadis, N. Arkani-Hamed, S. Dimopoulos, and G. Dvali,
        "New dimensions at a millimeter to a fermi and superstrings at a TeV",
        \emph{Physics Letters  B}, vol. 436, no. 3-4, pp. 257-263, 1998.

\bibitem{Hamed98} N. Arkani-Hamed, S. Dimopoulos, and G. Dvali, "The hierarchy problem and new dimensions at a millimeter",
         \emph{Physics Letters  B}, vol. 429, no. 3-4, pp. 263-272, 1998.

\bibitem{Hamed99} N. Arkani-Hamed, S. Dimopoulos, and G. Dvali,
        "Phenomenology, astrophysics and cosmology of theories with submillimeter dimensions and TeV scale quantum gravity",
        \emph{Physical Review D}, vol. 59, no. 5, Article ID 086004, 1999.

\bibitem{ACD} T. Appelquist, H. C. Cheng, and B. A. Dobrescu, "Bounds on universal extra dimensions",
        \emph{Physical Review D}, vol. 64, no. 3 , Article ID 035002, 2001.

\bibitem{Agashe} K. Agashe, N. G. Deshpande, and G. H. Wu,
        "Can extra dimensions accessible to the SM explain the recent measurement of anomalous magnetic moment of the muon?",
         \emph{Physics Letters  B}, vol. 511, no. 1, pp. 85-91, 2001.

 \bibitem{Agashe2} K. Agashe, N. G. Deshpande, and G. H. Wu,
         "Universal extra dimensions and $b\rar s \gamma$",
         \emph{Physics Letters  B}, vol. 514, no. 3-4, pp. 309-314, 2001.
%
\bibitem{Colangelo06} P. Colangelo, F. De Fazio, R. Ferrandes, and T. N. Pham,
        "Exclusive $B\rar K^{(*)} \ell^+ \ell^-$, $B\rar K^{(*)} \nu \bar{\nu} $  and $B\rar K^* \gamma $
        transitions in a scenario with a single universal extra dimension",
        \emph{Physical Review D}, vol. 73, no. 11 , Article ID 115006, 2006.

\bibitem{Haisch07} U. Haisch, and A. Weiler,
        "Bound on minimal universal extra dimensions from $ \bar{B} \rar X_s \gamma$",
        \emph{Physical Review D}, vol. 76, no. 3 , Article ID 034014, 2007.

\bibitem{Colangelo12} P. Biancofiore, P. Colangelo, and F. Fazio,
        "$B \rar K \eta ^{(')} \gamma $ decays in the standard model and in scenarios with universal extra dimensions",
        \emph{Physical Review D}, vol. 85, no. 9 , Article ID 094012, 2012.

\bibitem{Azizi12}   K. Azizi, S. Kartal, N. Kat{\i}rc{\i} , A. T. Olgun, and Z. Tavuko{\u{g}}lu,
        "Constraint on compactification scale via recently observed baryonic $\Lambda_b \rar \Lambda \ell^+ \ell^-$
        channel and analysis of the $\Sigma_b \rar \Sigma \ell^+ \ell^-$ transition in SM and UED scenario",
         \emph{Journal of High Energy Physics},  vol. 2012, article 24, 2012.

\bibitem{uoyilmaz12}U. O. Yilmaz,
        "Study of $B_c \rar D^*_s \ell^+ \ell^-$ in a Single Universal Extra Dimension",
        \emph{Physical Review D}, vol. 85, no. 11, Article ID 115026, 2012.


\bibitem {Belanger} G. Belanger, A. Belyaev, M. Brown, M. Kakizaki and A. Pukhov,
        "Testing minimal universal extra dimensions using Higgs boson searches at the LHC",
         \emph{Physical Review D}, vol. 87, no. 1 , Article ID 016008, 2013.


\bibitem {Flacke13} L. Edelhäuser, T. Flacke and  M. Krämer,
            "Constraints on models with universal extra dimensions from dilepton searches at the LHC"
            \emph{Journal of High Energy Physics},  vol. 2013, article 91, 2013.


\bibitem{Flacke14} T. Flacke, K. Kong and S. C. Park,
        "126 GeV Higgs in next-to-minimal Universal Extra Dimensions",
        \emph{Physics Letters  B}, vol. 728, pp. 262-267, 2014.

%
\bibitem{Buras03} A. J. Buras, M. Spranger, and A. Weiler,
        "The Impact of universal extra dimensions on the unitarity triangle and rare K and B decays",
        \emph{Nuclear Physics B}, vol. 660, no. 1-2, pp. 225-268, 2003.

\bibitem{Buras04} A. J. Buras, A. Poschenrieder, M. Spranger, and A. Weiler,
        "The impact of universal extra dimensions on $B\rar X_s \gamma$, $B\rar X_s \,gluon$, $B\rar X_s \mu^+ \mu^-$, $K_L \rar \pi^0 e^+ e^-$
        and $\varepsilon^{'}/\varepsilon$",
        \emph{Nuclear Physics B}, vol. 678, no. 1-2, pp. 455-490, 2004.

%
\bibitem{Colangelo06-2} P. Colangelo, F. De Fazio, R. Ferrandes, and T. N. Pham,
        "Spin effects in rare $B\rar X_s \tau^+ \tau^-$ and $K^{(*)} \tau^+ \tau^-$ decays in a single universal extra dimension scenario",
         \emph{Physical Review D}, vol. 74, no. 11, Article ID 115006, 2006.

\bibitem{Devidze06} G. Devidze, A. Liparteliani, and U. G. Meissner,
        "$B_{s,d}\rar \gamma \gamma$ decay in the model with one universal extra dimension",
        \emph{Physics Letters  B}, vol. 634, no. 1, pp. 59-62, 2006.


\bibitem{Mohanta07} R. Mohanta, and A. K. Giri,
        "Study of FCNC-mediated rare $B_s$  decays in a single universal extra dimension scenario",
        \emph{Physical Review D}, vol. 75, no. 3, Article ID 035008, 2007.


\bibitem{Colangelo08} P. Colangelo, F. De Fazio, R. Ferrandes, and T. N. Pham,
        "FCNC $B_s$ and $\Lambda_b$ transitions: Standard model versus a single universal extra dimension scenario",
        \emph{Physical Review D}, vol. 77, no. 5, Article ID 055019, 2008.


\bibitem{Saddique08} A. Saddique, M. J. Aslam, and C. D. Lu,
        "Lepton polarization asymmetry and forward–backward asymmetry in exclusive $B\rar K_1 \tau^+ \tau^-$ decay in universal extra dimension scenario",
        \emph{The European Physical Journal C}, vol. 56, no. 2, p. 267, 2008.


\bibitem{Aslam08} I. Ahmed, M. A. Paracha, and M. J. Aslam,
        "Exclusive $B\rar K(1) \ell^+ \ell^-$ decay in model with single universal extra dimension",
        \emph{The European Physical Journal C}, vol. 54, no. 4, p. 591, 2008.

\bibitem{Bashiry09} V. Bashiry, and K. Zeynali,
        "Exclusive $B\rar \pi \ell^+ \ell^- $ and $B\rar \rho \ell^+ \ell^- $ decays in the universal extra dimension,
         \emph{Physical Review D}, vol. 79, no. 3, Article ID 033006, 2009.


\bibitem{Colangelo09} M. V. Carlucci, P. Colangelo, and F. De Fazio,
        "Rare $B_s$ decays to $\eta$ and $\eta'$ final states",
        \emph{Physical Review D}, vol. 80, no. 5, Article ID 055023, 2009.


\bibitem{Li11}  Y. Li, and J. Hua,
        "Study of $B_s \rar \phi \ell^+ \ell^-$ decay in a single universal extra dimension model",
         \emph{The European Physical Journal C}, vol. 71, article. 1764, 2011.

%
\bibitem{Aliev07} T. M. Aliev, and M. Savci,
        "$\Lambda_b \rar \Lambda \ell^+ \ell^-$ decay in universal extra dimensions",
        \emph{The European Physical Journal C}, vol. 50, no. 1, p. 91, 2007.


\bibitem{Azizi11} N. Katirci, and K. Azizi,
        "Investigation of the $\Lambda_b \rar \Lambda \ell^+ \ell^-$ transition in universal extra dimension using form factors from full QCD",
        \emph{Journal of High Energy Physics},  vol. 2011, article 87, 2011.


\bibitem{Azizi11-2} N. Katirci and K. Azizi,
        "B to strange tensor meson transition in a model with one universal extra dimension",
        \emph{Journal of High Energy Physics},  vol. 2011, article 43, 2011.
%
%
\bibitem{Geng2002} C. Q. Geng, C. W. Hwang, and C. C. Liu,
        "Study of rare ${B}_{c}^{+}\ensuremath{\rightarrow}{D}_{d,s}^{(*)+}l\overline{l}$ decays",
        \emph{Physical Review D}, vol. 65, no. 9, Article ID 094037, 2002.


\bibitem{Faessler} A. Faessler, Th. Gutsche, M. A. Ivanov, J. G. K\"{o}rner, and V. E. Lyubovitskij,
        "The exclusive rare decays  $B\rar k \bar \ell \ell$ and  $B_c\rar D(D^*)\bar\ell \ell$ in a relativistic quark model",
        \emph{The European Physical Journal C  direct}, vol. 4, no. 1, p. 1, 2002.

\bibitem{Ho10}     H.-M. Choi,
        "Light-front quark model analysis of the exclusive rare $B_c \rightarrow  D_{(s)} (\ell^{+} \ell^-, \nu_{\ell} \overline{\nu}_{\ell})$ decays",
         \emph{Physical Review D}, vol. 81, no. 5, Article ID 054003, 2010.

\bibitem{Ebert10}  D. Ebert, R. N. Faustov, and V. O. Galkin,
        "Semileptonic and nonleptonic decays of ${B}_{c}$ mesons to orbitally excited heavy mesons in the relativistic quark model",
         \emph{Physical Review D}, vol. 82, no. 3, Article ID 034032, 2010.


\bibitem{Azizi08}  K. Azizi, and R. Khosravi,
        "Analysis of the rare semileptonic $B_c \rar P(D, D_s)\ell^+ \ell^- / \nu \bar\nu$ decays within QCD sum rules",
        \emph{Physical Review D}, vol. 78, no. 3, Article ID 036005, 2008.


%
\bibitem{Buras96}  G. Buchalla, A. J. Buras, and M. Lautenbacher,
        "Weak decays beyond leading logarithms",
        \emph{Reviews of Modern Physics}, vol. 68, no. 4, p. 1125, 1996
%
\bibitem{Buras95}  A. J. Buras, and M. M\"{u}nz,
        "Effective Hamiltonian for $B \rar X_s e^+ e^-$ beyond leading logarithms in the naive dimensional
                            regularization and 't Hooft-Veltman schemes",
        \emph{Physical Review D}, vol. 52, no. 1, p. 186, 1995.

%
\bibitem{Misiak93} M. Misiak,
        "The $b \rar s e^+ e^-$ and $b \rar s \gamma$ decays with next-to-leading logarithmic QCD-corrections",
        \emph{Nuclear Physics B}, vol. 393, no. 1-2, pp. 23-45, 1993. \emph{Erratum:} vol. 439, no. 1-2, pp. 461-465, 1995.

\bibitem{AAli91}  A. Ali, T. Mannel, and T. Morozumi,
        "Forward-backward asymmetry of dilepton angular distribution in the decay $b \rar s \ell^+ \ell^-$",
        \emph{Physics Letters  B}, vol. 273, no. 4, pp. 505-512, 1991.

\bibitem{pdg} K. Nakamura, {\it et al.} (Particle Data Group),
    "Review of Particle Physics",
    \emph{Journal of Physics G: Nuclear and Particle Physics}, vol. 37, Article No: 075021, 2010.

%
\bibitem {AAli2000} A. Ali, P. Ball, L. T. Handoko, and G. Hiller,
        "Comparative study of the decays $B \rar (K, K^*) \ell^+ \ell^-$ in the standard model and supersymmetric theories",
        \emph{Physical Review D}, vol. 61, no. 7, Article ID 074024, 2000.

\bibitem{Fukae2000} S. Fukae, C. S. Kim, and T. Yoshikawa,
        "Systematic analysis of the lepton polarization asymmetries in the rare B decay $B \rar X_s \tau^+ \tau^-$",
        \emph{Physical Review D}, vol. 61, no. 7, Article ID 074015, 2000.

\bibitem{Kruger96}  F. Kr\"{u}ger, and L. M. Sehgal,
        "Lepton polarization in the decays $B \rar X_s \mu^+ \mu^-$ and $B \rar X_s \tau^+ \tau^-$,
        \emph{Physics Letters  B}, vol. 380, no. 1-2, pp. 199-204, 1996.

%
\end{thebibliography}
\end{document}